\documentclass[12pt]{article}
\begin{document}

\title{Regge calculus from discontinuous metrics}
\author{V.M.Khatsymovsky \\
 {\em Budker Institute of Nuclear Physics} \\ {\em
 Novosibirsk,
 630090,
 Russia}
\\ {\em E-mail address: khatsym@inp.nsk.su}}
\date{}
\maketitle
\begin{abstract}
Regge calculus is considered as a particular case of the
more general system where the linklengths of any two
neighbouring 4-tetrahedra do not necessarily coincide on
their common face. This system is treated as that one
described by metric discontinuous on the faces. In the
superspace of all discontinuous metrics the Regge calculus
metrics form some hypersurface. Quantum theory of the
discontinuous metric system is assumed to be fixed somehow
in the form of quantum measure on (the space of functionals
on) the superspace. The problem of reducing this measure
to the Regge hypersurface is addressed. The quantum Regge
calculus measure is defined from a discontinuous metric
measure by inserting the $\delta$-function-like phase
factor. The requirement that this reduction would respect
natural physical properties (positivity, well-defined
continuum limit, absence of lattice artefacts) put rather
severe restrictions and allows to define practically
uniquely this phase factor.
\end{abstract}
\newpage
Since the idea had been put forward that Regge calculus
should be formulated in terms of the areas of the triangles
rather than the edge lengths \cite{Rov} the so-called area
Regge calculus model was of a certain
interest \cite{BarRocWil,RegWil}. In this model the areas
of the 2-faces (triangles) are treated as independent
variables. Thereby configuration space of the theory
becomes larger, and ordinary Regge calculus corresponds
only to some subset (hypersurface) in this space. Despite
of this, the equations of motion of the theory get
simplified. In particular, a kind of the area Regge
calculus, area tensor-connection one \cite{Kha} possesses
the set of commuting constraints (in the continuous time
limit) and can be quantised resulting in the finite
expectation values for areas \cite{Kha1}. The problem is
how to pass from areas to lengths. The constraints
enforcing areas to be expressible in terms of the edge
lengths have been discussed in \cite{Mak,MakWil}. Now the
problem is dynamical one, e. g. given the Feynman path
integral on the configuration space of area Regge calculus,
how could we define such an integral on the subspace
corresponding to the usual length-based Regge calculus?

In the area Regge calculus the two 4-simplices $\sigma^4$
sharing a given 3-face $\sigma^3$ do not necessarily have
coinciding linklengths on this face, only the 4 conditions
equating the areas of the triangles $\sigma^2$ on this
3-face are required. One can go further and discuss the
4-simplices with completely independent lengths, i. e.
impose no conditions at all. Area Regge calculus will be a
particular case of this configuration. Introducing metric
tensor $g_{\lambda\mu}$ one can say that it is
discontinuous on the 3-faces. Regge calculus is then
specified by setting continuity conditions on the induced
3-face metric $g^{\|}_{ab}$ on the 3-faces $\sigma^3$,
\begin{equation}                                         
\label{delta-g}
\Delta_{\sigma^3}g^{\|}_{ab}\stackrel{\rm def}{=}
g^{\|}_{ab}(\sigma^4_1)-g^{\|}_{ab}(\sigma^4_2)=0
\end{equation}

\noindent where $\sigma^4_1$, $\sigma^4_2$ are the two
4-simplices sharing the 3-face $\sigma^3$ = $\sigma^4_1\cap
\sigma^4_2$. (At the same time the normal-to-face metric
$g^{\bot}$ is discontinuous thus providing nonzero angle
defects).

The conditions (\ref{delta-g}) define some hypersurface
$\Gamma_{\rm cont}$ in the superspace of simplicial
(piecewise-flat) metrics including discontinuous ones.
Suppose quantum theory on this superspace is fixed in the
form of the Feynman path integral. That is, a quantum
measure is given. The measure can be viewed as a linear
functional $\mu (\Psi)$ on the space of the functionals
$\Psi (\{g\})$ on the superspace. To pass to the quantum
theory on the hypersurface $\Gamma_{\rm cont}$ we use some
analogy with the quantum-mechanical notion of the "state",
although full analogy is absent since we do not mean
decomposing the spacetime into space and time. In
our case the functionals $\Psi (\{g\})$ are intended to
describe the states. (Rather these could be considered as
analogs of quantum-mechanical $\Phi^*\Phi$.) We are
interested in the particular case of these with support on
$\Gamma_{\rm cont}$ of the form
\begin{equation}                                         
\Psi (\{g\})=\psi (\{g\})\delta_{\rm cont}(\{g\})
\end{equation}

\noindent where $\delta_{\rm cont}(\{g\})$ is
(many-dimensional) $\delta$-function with support on
$\Gamma_{\rm cont}$. Generally derivatives of
$\delta$-function also ensure given support, but these
would violate positivity in what follows. If the measure
can be defined on such the functionals then we define the
measure on $\Gamma_{\rm cont}$,
\begin{equation}                                         
\mu_{\rm cont}(\cdot )
=\mu (\delta_{\rm cont}(\{g\})~\cdot ).
\end{equation}

To construct $\delta_{\rm cont}$ first reveal irreducible
set of the constraints on the metric of the type of
(\ref{delta-g}) in terms of linklengths. For that consider
all the 4-simplices containing the given link, i. e. the
star of the link. The surface of the star is topologically
the 2-sphere and is depicted in the Fig. 1.

\begin{figure}
\label{star}
\begin{picture}(16,16)(-7,0)
\thinlines
\put(3,12){\line(4,1){4}}
\put(6,14){\line(1,-1){1}}
\put(7,13){\line(2,1){2}}
\put(7,13){\line(5,-1){5}}

\put(3,12){\line(2,-1){2}}
\put(7,13){\line(2,-3){2}}
\put(2,9){\line(3,2){3}}
\put(4,8){\line(1,3){1}}

\put(4,8){\line(5,2){5}}
\put(8,7){\line(1,3){1}}
\put(9,10){\line(3,-2){3}}
\put(9,10){\line(5,-1){5}}

\put(12,12){\line(2,-3){2}}
\put(2,7){\line(0,1){2}}
\put(2,7){\line(2,1){2}}
\put(3,6){\line(1,2){1}}

\put(4,8){\line(1,-1){3}}
\put(7,5){\line(1,2){1}}
\put(8,7){\line(1,-1){2}}
\put(10,5){\line(2,3){2}}

\put(12,8){\line(1,-2){1}}
\put(13,6){\line(1,3){1}}
\put(2,7){\line(1,-3){1}}
\put(3,4){\line(0,1){2}}

\put(3,6){\line(3,-4){3}}
\put(6,2){\line(1,3){1}}
\put(7,5){\line(1,-3){1}}
\put(7,5){\line(2,-1){4}}

\put(10,5){\line(1,-2){1}}
\put(11,3){\line(2,3){2}}
\multiput(7,8)(1,1){2}{\line(1,1){0.75}}
\multiput(7,13)(0,-1.35){4}{\line(0,-1){0.9}}
\multiput(5,11)(0.7,-1.05){3}{\line(2,-3){0.5}}
\thicklines
\put(3,12){\line(3,2){3}}
\put(6,14){\line(1,0){3}}
\put(9,14){\line(3,-2){3}}
\put(5,11){\line(1,1){2}}

\put(5.075,10.925){\line(1,1){2}}
\put(9,10){\line(3,2){3}}
\put(2,9){\line(1,3){1}}
\put(5,11){\line(4,-1){4}}

\put(2,9){\line(2,-1){2}}
\put(4,8){\line(4,-1){4}}
\put(8,7){\line(4,1){4}}
\put(12,8){\line(2,1){2}}

\put(2,7){\line(1,-1){1}}
\put(3,6){\line(4,-1){4}}
\put(7,5){\line(1,0){3}}
\put(10,5){\line(3,1){3}}

\put(3,4){\line(3,-2){3}}
\put(6,2){\line(1,0){2}}
\put(8,2){\line(3,1){3}}

\put(7.6,12.35){\scriptsize A}
\put(4.75,11.25){\scriptsize B}
\put(8.95,9.3){\scriptsize C}
\put(12.1,12.1){\scriptsize D}
\put(9,14.1){\scriptsize E}
\put(5.6,14.1){\scriptsize F}
\put(2.6,12.1){\scriptsize G}
\put(1.5,9){\scriptsize H}
\put(4,7.45){\scriptsize I}
\put(7.75,7.2){\scriptsize J}
\put(11.8,8.2){\scriptsize K}
\put(14.1,9.1){\scriptsize L}
\put(1.5,7){\scriptsize P}
\put(3.3,6.1){\scriptsize Q}
\put(6.35,4.55){\scriptsize R}
\put(13.05,5.45){\scriptsize T}
\put(9.8,5.25){\scriptsize S}
\put(2.55,3.55){\scriptsize V}
\put(5.7,1.45){\scriptsize W}
\put(8,1.45){\scriptsize X}
\put(11,2.5){\scriptsize Y}
\put(6.5,7.6){\scriptsize O}

\put(8.3,11.7){$s_{\rm O(ACD)}$}
\put(5.4,6.65){$s_{\rm O(IJR)}$}
\put(10.3,10.2){$s_{\rm O(CDL)}$}
\put(8.6,8.1){$s_{\rm O(CJK)}$}
\put(8.6,6.6){$s_{\rm O(JKS)}$}
\put(5,4){\dots}

\end{picture}
\caption{The star of the link.}
\end{figure}
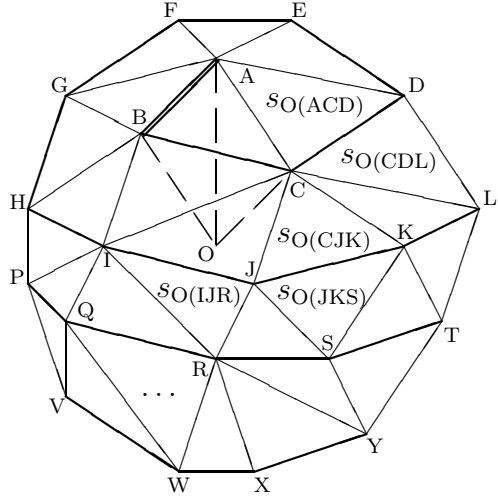

In this picture the $k$-simplices $\sigma^k$ mean the 4D
$k+1$-simplices ${\rm O}'\sigma^k$, in particular, the
considered link ${\rm O}'{\rm O}$ is represented here by
the vertex O, and it's length squared defined in the
4-simplex, say, ${\rm O}'{\rm OABC}$ is denoted here by
$s_{\rm O(ABC)}$. The fact that these values are the same
in all the 4-simplices ${\rm O}'\sigma^k$ is taken into
account by the product of $\delta$-functions over all the
links on the 2-sphere Fig.1,
\begin{equation}                                         
\label{delta2}
\prod_{\sigma^2\supset {\rm O}}{\delta (\Delta_{\sigma^2}s
_{\rm O})}
\end{equation}

\noindent where $\Delta_{\rm OAB}s_{\rm O}$ = $s_{\rm
O(ABC)}$ - $s_{\rm O(ABG)}$, etc. This product containes
redundant $\delta$-functions connected with occurence of
cycles enclosing separate links $\sigma^1$ = OA, OB, \dots
(corresponding to vertices on the 2-sphere Fig.1). To
exclude cycles, draw some (2D in our picture) cut
$H^{(2)}_{\rm O}$ in the star of the point O consisting of
the triangles and passing through each radial link. By
definition, $H^{(2)}_{\rm O}$ is a collection of the
triangles on which continuity conditions for $s_{\rm O}$
should not be imposed. This cut should be chosen in such a
way that it, first, would not consist of components
intersecting only at the point O (in order that cycles in
the compliment of the cut in the star were absent indeed),
and, second, would not divide interior of the star into the
disconnected components (otherwise $s_{\rm O}$ in these
components would be allowed to be different). In other
words, intersection of $H^{(2)}_{\rm O}$ with the 2-sphere
$S^{(2)}_{\rm O}$ of Fig.1 should be (poligonal) continuous
line without self-intersections.

To cancel redundant $\delta$'s in (\ref{delta2}) one
should divide (\ref{delta2}) by many-dimensional
$\delta$-function which would express continuity of $s_{\rm
O}$ on $H^{(2)}_{\rm O}$. A sufficient condition for such
the continuity is expressed by the product of
$\delta$-functions over all the vertices on the 2-sphere
Fig.1,
\begin{equation}                                         
\label{delta1}
\prod_{\sigma^1\supset {\rm O}}{\delta (\Delta_{\sigma^1}s
_{\rm O})}.
\end{equation}

\noindent  Here $\Delta_{\sigma^1}s_{\rm O}$ $\stackrel{\rm
def}{=}$ $\Delta_{\sigma^2(\sigma^1)}s_{\rm O}$ where
$\sigma^2(\sigma^1)$ is a one of the two triangles in
$H^{(2)}_{\rm O}$ containing $\sigma^1$, $H^{(2)}_{\rm O}$
$\supset$ $\sigma^2(\sigma^1)$ $\supset$ $\sigma^1$.

In turn, (\ref{delta1}) containes redundant
$\delta$-functions (more accurately, one such function).
Indeed, the number of edges in the continuous line without
intersections $H^{(2)}_{\rm O}$ $\cap$ $S^{(2)}_{\rm O}$ is
by one smaller than the number of vertices through which
the line passes. This means that $\sigma^2(\sigma^1)$
involved in the definition of (\ref{delta1}) above is not
one-to-one correspondence, and there is a pair of the two
neighbouring radial links, $\sigma^1_1$, $\sigma^1_2$, the
edges of the same triangle $\sigma^2(\sigma^1_1)$ = $\sigma
^2(\sigma^1_2)$ so that the $\delta$-function of $\Delta
_{\sigma^1_1}s_{\rm O}$ = $\Delta_{\sigma^1_2}s_{\rm O}$
enters (\ref{delta1}) twice. Therefore one should divide
(\ref{delta1}) by
\begin{equation}                                         
\label{delta0}
\delta (\Delta_{\rm O}s_{\rm O})
\end{equation}

\noindent where $\Delta_{\rm O}s_{\rm O}$ $\stackrel{\rm
def}{=}$ $\Delta_{\sigma^1_1}s_{\rm O}$ = $\Delta_{
\sigma^1_2}s_{\rm O}$. Again situation can be described as
occurence of some reminder cycle, now in the cut $H^{(2)}
_{\rm O}$ itself treated in some generalised way taking
into account the number of times the cut passes the same
triangle, and (\ref{delta0}) just serves to cancel effect
of this cycle. Such the cut OBABCDE\dots passing OAB twice
is just shown in Fig.1.

The resulting $\delta$-function stating unambiguity of
$s_{\rm O}$ looks like
\begin{equation}                                         
\label{delta2-1+0}
\prod_{\sigma^2\supset {\rm O}}{\delta (\Delta_{\sigma^2}s
_{\rm O})}\left(\prod_{\sigma^1\supset {\rm O}}{\delta
(\Delta_{\sigma^1}s_{\rm O})}\right)^{-1}\delta
(\Delta_{\rm O}s_{\rm O}).
\end{equation}

\noindent Here the number of $\delta$-functions is $L$
$-$ $P$ + 1 where $L$, $P$ are the numbers of the links and
vertices on the 2-sphere Fig.1, respectively. In some
functional integral these are to be integrated over
\begin{equation}                                         
{\rm d}s_{\rm O(ABC)}{\rm d}s_{\rm O(ACD)}{\rm d}s_{\rm
O(ADE)}\dots,
\end{equation}

\noindent the overall number of integrations being $T$, the
number of the triangles on the 2-sphere Fig.1. As a result,
we are left with $T$ $-$ $L$ + $P$ $-$ 1 = 1 integration as
it follows from the Euler characteristic 2 for the
2-sphere.

Upon restoring the 4D notations the eq. (\ref{delta2-1+0})
reads
\begin{equation}                                         
\label{delta3-2+1}
\prod_{\sigma^3\supset {\rm O'O}}{\delta (\Delta_{\sigma^3}
s_{\rm O'O})}\left(\prod_{\sigma^2\supset {\rm O'O}}{\delta
(\Delta_{\sigma^2}s_{\rm O'O})}\right)^{-1}\delta
(\Delta_{\rm O'O}s_{\rm O'O}).
\end{equation}

\noindent Take the product of these expressions over all
the links ${\rm O}'{\rm O}$. Rearranging, we have for
separate factors
\begin{equation}                                        
\prod_{\sigma^1}{\prod_{\sigma^k\supset\sigma^1}{\delta
(\Delta_{\sigma^k}s_{\sigma^1})}} = \prod_{\sigma^k}{\prod
_{\sigma^1\subset\sigma^k}{\delta(\Delta_{\sigma^k}s
_{\sigma^1})}} = \prod_{\sigma^k}{\delta^{k{k+1\over 2}}
(\Delta_{\sigma^k}S_{\sigma^k})}
\end{equation}

\noindent thus expressing these in terms of the
discontinuities of the edge component metric $S_{\sigma^k}$
\cite{PirWil} induced on the
$k$-simplex and being simply collection of $k{k+1\over 2}$
it's edge lengths squared. The total product thus reads
\begin{equation}                                        
\label{delta3-2+1'}
\prod_{\sigma^3}{\delta^6(\Delta_{\sigma^3}S_{\sigma^3})}
\left (\prod_{\sigma^2}{\delta^3(\Delta_{\sigma^2}S_{\sigma
^2})}\right )^{-1}\prod_{\sigma^1}{\delta (\Delta_{\sigma
^1}S_{\sigma^1})}.
\end{equation}

\noindent Some remark should be made concerning the middle
factor in this expression where the metric discontinuity
$\Delta_{\sigma^2}S_{\sigma^2}$ for the different
components of metric $S_{\sigma^2}$, in principle, may turn
out to be defined between the {\it different} pairs of the
neighbouring 4-simplices containing the given triangle
$\sigma^2$, if one uses the above construction. Namely, for
each $\sigma^2$ containing the given $\sigma^1$ we take
$\sigma^3$, one of the two 3-faces containing $\sigma^2$
and belonging to the 3D cut $H^{(3)}_{\sigma^1}$ in the
star of the link $\sigma^1$ (the above $H^{(2)}_{\rm O}$ in
3D language). On this $\sigma^3$ the considered
discontinuity
of the $\sigma^2$-metric just should be taken. However, as
we see, the choice of this $\sigma^3$ depends (by means of
$H^{(3)}_{\sigma^1}$) on the edge $\sigma^1$ of this
triangle, i.e. may be different for the different
components. Fortunately, it is possible to change the
choice of this 3-face not violating the result so that it
would become the same for the edges of a given triangle.
Indeed, singularity connected with the cycle enclosing the
triangle is contained in the product of the type $\delta
(s_1-s_2)\delta (s_2-s_3)\dots\delta (s_N-s_1)$ over all
$N$ 3-faces containing the triangle and can be cancelled by
dividing it by any of the factors yielding the same result.
Thereby definition of the discontinuity of the metric
induced on the triangle can be adjusted to be the same for
all the metric components.

The form (\ref{delta3-2+1'}) looks suitable for setting
requirements which would help to fix a factor multiplying
the $\delta$ functions in the measure. A natural assumption
is that the result of implementing continuity condition for
the induced $k$-metric across a $k$-face should not depend
on the real form and size of the $k$-face, only on the
hyperplane spanned by this face. This assumption looks
natural if we consider Regge links as imaginary objects
only used for the needs of triangulation on some smooth
background metric. Of course,
something like "interaction" of the Regge links may exist,
but this is supposed to be introduced by "physical" part of
the theory by which we mean the theory of discontinuous
metric distributions already fixed somehow in the form of
the functional integral, while what do we consider here is
a kind of kinematical phase factor in which we are trying
to minimize any dynamical effect especially introduced "by
hand". It is convenient to rewrite corresponding $\delta$
function in terms of metric tensor $g_{\lambda\mu}$ and of
a set of $k$ 4-vectors $\tau^\lambda_a$ defining the
$k$-face
and the induced metric on it, $g^{\|}_{ab}$ = $\tau^\lambda
_a\tau^\mu_bg_{\lambda\mu}$. Then it is clear that the
required factor is determinant of $g^{\|}_{ab}$ raised to
the power ${1\over 2}(k+1)$, i. e. the $(k+1)$-th power of
the $k$-face volume $V_{\sigma^k}$. The resulting $\delta$
factor reads
\begin{equation}                                        
\label{inv-delta}
[{\rm det}(\tau^\lambda_a\tau^\mu_bg_{\lambda\mu})]^{k+1
\over 2}\delta^{k{k+1\over 2}}(\tau^\lambda_a\tau^\mu_b
\Delta_{\sigma^k}g_{\lambda\mu}) = V^{k+1}_{\sigma^k}
\delta^{k{k+1\over 2}}(\Delta_{\sigma^k}S_{\sigma^k}).
\end{equation}

\noindent It is invariant w.r.t. the deformations of the
$k$-face keeping it placed in a fixed $k$-plane (the metric
$g_{\lambda\mu}$ being fixed) $\tau^\lambda_a$ $\mapsto$
$m^b_a\tau^\lambda_b$ with arbitrary $k$ $\times$ $k$
matrix $m$. This property looks natural also from
the viewpoint of existence of the continuum limit and, in
particular, of absence of the lattice artefacts if we do
not want to introduce these artefacts "by hand".

The final answer reads
\begin{eqnarray}                                        
\label{delta-cont}
\delta_{\rm cont}&=&\prod_{\sigma^3}{V^4_{\sigma^3}\delta^6
(\Delta_{\sigma^3}S_{\sigma^3})}\left (\prod_{\sigma^2}
{V^3_{\sigma^2}\delta^3(\Delta_{\sigma^2}S_{\sigma^2})}
\right )^{-1}\prod_{\sigma^1}{V^2_{\sigma^1}\delta
(\Delta_{\sigma^1}S_{\sigma^1})}\nonumber\\&=&
\prod_{\sigma^3}{V^4_{\sigma^3}}
\prod_{\sigma^2}{V^{-3}_{\sigma^2}}
\prod_{\sigma^1}{V^2_{\sigma^1}}
\prod_{\sigma^1,~\sigma^3\supset\sigma^1}
\hspace{-5mm}^{\prime}\hspace{5mm}
{\delta (\Delta_{\sigma^3}s_{\sigma^1})}
\end{eqnarray}

\noindent where the prime on the product means that the
redundant $\delta$'s are omitted.

Consider consequences of taking a more simple expression
for $\delta_{\rm cont}$. For example, take in
(\ref{delta-cont}) only the primed product of
$\delta$-functions and only slightly change it according to
the recipe (\ref{inv-delta}) at $k$ = 1 considering
argument of each $\delta$-function there as discontinuity
of the induced 1-metric (the length squared) across the
1-face (the link). Correspondingly, multiply each function
by $s_{\sigma^1}$. Consider the star of a given link
${\rm O}'{\rm O}$, i. e. Fig. 1 in 3D language. The eq.
(\ref{delta3-2+1}) turns out to be multiplied by $(T-1)$-th
power of $s_{\rm O'O}$, $T$ being the number of the
4-simplices containing this link. Upon integrating out the
$\delta$-functions we are left with the factor
\begin{equation}                                        
\label{naive}
s_{\rm O'O}^{T-1}{\rm d}s_{\rm O'O}
\end{equation}

\noindent in the measure corresponding to the
$s_{\rm O'O}$-dependence. Suppose the partial integration
in the functional integral over ${\rm d}s_{\rm O'O}$ is
made, and we are interested in the dependence on the
distance between any two neighbouring vertices, say A and
B, in the star of our link. Confluence of these two
vertices together will lead to changing the index in
(\ref{naive}) by $-2$. That is, there is discontinuity at
the point $s_{\rm AB}$= 0. This contradicts to our
assumption on Regge links aspurely triangulation objects on
some smooth background metricsince then tending A to B
would be smooth process.

This "no-go" example advises to view our result
(\ref{delta-cont}) from a more general position. The
issuing point are already discussed $\delta$-factors of the
type of (\ref{inv-delta}) invariant w.r.t. changing the
form of the $k$-face. For a given $k$-face $\sigma^k$ these
factors can differ by the choice $i$ = 1, \dots, $n_k$ of
the 3-face $\sigma^3$ = $\sigma^3_i(\sigma^k)$ $\supset$
$\sigma^k$ across which discontinuity of the metric is
defined. As we have discussed just now, the number of such
factors per $k$-face $n_k$ should be fixed not depending on
the neighbourhood of the $k$-face, and general form of
$i$-th factor for all the $k$-faces looks like
\begin{equation}                                        
\delta_{ik}=\prod_{\sigma^k}{V^{k+1}_{\sigma^k}\delta^{k
{k+1\over 2}}(\Delta_{\sigma^3_i(\sigma^k)}S_{\sigma^k})},
\end{equation}

\noindent $i$ = 1, \dots, $n_k$, $n_3$ = 1 of which the
overall factor is constructed,
\begin{equation}                                        
\delta_{\rm cont}=\delta^{\varepsilon_{13}}_{13}\prod^{n_2}
_{i=1}\delta^{\varepsilon_{12}}_{12}\prod^{n_1}_{i=1}\delta
^{\varepsilon_{i1}}_{i1}
\end{equation}

\noindent where $\varepsilon_{ik}$ are some integers. These
integers being larger than 1 or negative may have sense if,
as above, upon formal cancellation of the $\delta$'s in the
numerator and denominator we come to the well defined
expression which in this case should take the form
\begin{equation}                                        
\delta_{\rm cont}=\left (\prod_{\sigma^3}{V^4_{\sigma^3}}
\right )^{\varepsilon_{13}}\left (\prod_{\sigma^2}{V^3
_{\sigma^2}}\right )^{\sum^{n_2}_{i=1}{\varepsilon_{i2}}}
\left (\prod_{\sigma^1}{V^2_{\sigma^1}}\right )^{\sum^{n_1}
_{i=1}{\varepsilon_{i1}}}
\prod_{\sigma^1,~\sigma^3\supset\sigma^1}
\hspace{-5mm}^{\prime}\hspace{5mm}
{\delta (\Delta_{\sigma^3}s_{\sigma^1})}.
\end{equation}

\noindent Original number of the length variables
associated to the link is the number of the 4-simplices in
the star of the link or the number of the triangles $T$ on
the 2-sphere Fig. 1. The final number of the variables with
taking into account $\delta$-functions should be 1,
\begin{equation}                                        
T-L\varepsilon_{13}-P\sum^{n_2}_{i=1}{\varepsilon_{i2}}
-\sum^{n_1}_{i=1}{\varepsilon_{i1}}=1.
\end{equation}

\noindent Comparing this to the Euler characteristic
equation for the 2-sphere we return to the eq.
(\ref{delta-cont}).

This discussion admits extension to arbitrary spacetime
dimensionality $d$, here without rigorous topological
proof, however. The formula (\ref{delta-cont}) is naturally
generalised to
\begin{equation}                                        
\label{delta-cont-d}
\delta^{(d)}_{\rm cont}=\delta^{(d)}_{d-1}(\delta^{(d)}
_{d-2})^{-1}\delta^{(d)}_{d-3}\dots (\delta^{(d)}_1)
^{(-1)^d}
\end{equation}

\noindent where
\begin{equation}                                        
\delta^{(d)}_k=\prod_{\sigma^k}{V^{k+1}_{\sigma^k}\delta
^{k{k+1\over 2}}(\Delta_{\sigma^{d-1}(\sigma^k)}
S_{\sigma^k})}.
\end{equation}

\noindent This form ensures the required number 1 of the
independent length variables per link upon taking into
account the $\delta$-functions. Indeed, let $N^{(d)}
_k(\sigma^1)$ denotes the number of $k$-simplices
containing the link $\sigma^1$. Then the considered number
of the independent length variables being read off from
(\ref{delta-cont-d}) is
\begin{equation}                                        
\label{Euler-d}
N^{(d)}_d(\sigma^1)-N^{(d)}_{d-1}(\sigma^1)+\dots
+(-1)^dN^{(d)}_2(\sigma^1)+(-1)^{d+1}.
\end{equation}

\noindent At the same time $N^{(d)}_k(\sigma^1)$ can be
viewed as the number of $(k-2)$-simplices on the surface of
the star of the link $\sigma^1$ which possesses topology of
the $(d-2)$-sphere, and (\ref{Euler-d}) is just topology
invariant, namely, Euler characteristic for the
$(d-2)$-sphere minus 1 \cite{MilSta}, i. e. just 1.

Thus, the assumption that in quantum theory Regge calculus
is a kind of state of a more general quantised system where
discontinuous metrics are allowed leads, in the natural
physical assumptions of positivity and nonsingularity in
the continuum limit, to the unique version of the Regge
calculus quantum measure. Universality of the result is
also seen from extendability to the arbitrary spacetime
dimensionality and from the deep connection with the
(local) topological properties of the spacetime.

The result obtained can be applied to the area Regge
calculus. (The quantum measure in this theory can be
identically rewritten by inserting some $\delta$-functions
as the measure in the theory of discontinuous metrics.)
A particular feature of the phase factor
$\delta_{\rm cont}$ is that it is invariant w.r.t. the
overall rescaling linklengths. Therefore one can think that
cutoff properties of the original area Regge calculus
measure at small or large distances hold also for the usual
(length-based) Regge calculus measure obtained with the
help of the present construction, and finite expectation
values of areas \cite{Kha1} would mean also finite
expectation values of the linklengths \cite{Kha2}.

Another application is to the ordinary Regge calculus
action,
\begin{equation}                                        
\label{Regge}
\sum_{\sigma^2}{V_{\sigma^2}\varphi_{\sigma^2}}.
\end{equation}

\noindent Here angle defect $\varphi_{\sigma^2}$ is defined
by the dihedral angles of separate 4-simplices $\sigma^4$
containing $\sigma^2$ and naturally extends to the case of
discontinuous metric when separate 4-simplices become
completely independent. Also the area $V_{\sigma^2}$ can be
treated as area of any of the 4-simplices containing
$\sigma^2$. Thus, (\ref{Regge}) can be extended in many
ways to the discontinuous metric so that it becomes a sum
of independent terms referred to separate 4-simplices. It
is natural to write down quantum measure for such the
ultralocal system as the product of the measures for
separate 4-simplices. Since there is one-to-one
correspondence between the number of the linklengths of a
4-simplex $\sigma^4$ and the number of the components of
the metric $g_{\lambda\mu}$ = $const$ inside $\sigma^4$,
there is also correspondence between possible 4-simplex
measures and possible local measures at a point in the
continuum theory. Thereby quantisation of the action
(\ref{Regge}) in the framework of our approach is formally
defined, and the question is whether contours of
integration in the path integral exist such that it would
converge as in the quantised area Regge calculus model
mentioned above.

\bigskip

The present work was supported in part by the Russian
Foundation for Basic Research through Grant No.
03-02-17612, through Grant No. 00-15-96811 for Leading
Scientific Schools and by the Ministry of Education Grant
No. E00-3.3-148.

\end{document}